\documentclass[]{interact}

\usepackage{epstopdf}% To incorporate .eps illustrations using PDFLaTeX, etc.
\usepackage[caption=false]{subfig}% Support for small, `sub' figures and tables

\usepackage[numbers,sort&compress]{natbib}% Citation support using natbib.sty
\bibpunct[, ]{[}{]}{,}{n}{,}{,}% Citation support using natbib.sty
\renewcommand\bibfont{\fontsize{10}{12}\selectfont}% Bibliography support using natbib.sty
\makeatletter% @ becomes a letter
\def\NAT@def@citea{\def\@citea{\NAT@separator}}% Suppress spaces between citations using natbib.sty
\makeatother% @ becomes a symbol again

\theoremstyle{plain}% Theorem-like structures provided by amsthm.sty

\theoremstyle{definition}

\theoremstyle{remark}

\begin{document}

%\articletype{ARTICLE TEMPLATE}% Specify the article type or omit as appropriate

\title{Light propagation and observed light propagation}

\author{
\name{Wei Guo\textsuperscript{a}\thanks{CONTACT W. Guo. Email: weiguoguo@hotmail.com}}
\affil{\textsuperscript{a}P. O. Box 470011, Charlotte, North Carolina 28247, USA}
}

\maketitle

\begin{abstract}
Light propagation is viewed as a process involving mutual creation of electric and magnetic fields.  This viewpoint is used to argue that the conventional retarded solutions to electromagnetic wave equations (whose source is a current density in this work) are wrong, and to solve the wave equations in a different way by tacking every step in the said process.  It turns out that the solutions to the wave equations, or the emitted fields from the current density, have equally weighted advanced and retarded components.  After these components are explained for their mathematical and physical origins, it is then pointed out that the emitted fields are related but not identical to those fields observed in light observation.  The latter fields are calculated and found, as expected, to be retarded.
\end{abstract}

%\begin{keywords}
%Sections; lists; figures; tables; mathematics; fonts; %references; appendices
%\end{keywords}

\section{Introduction}
Usually, light is viewed as an electromagnetic wave---see Ref. \cite{kle:86} for other viewpoints---for the reason that, in vacuum, the emitted electric field $\vec{E}(\vec{r},t)$ and magnetic field $\vec{B}(\vec{r},t)$ from a current density $\vec{j}(\vec{r},t)$, all evaluated at position $\vec{r}$ and time $t$, each satisfy a wave equation, in which the common wave propagation speed $c$ is nothing other than the speed of light in vacuum: 
\begin{equation}
\label{e1}
\nabla ^{2}\vec{E}-\frac{1}{c^{2}}\frac{\partial ^{2}}{\partial t^{2}}\vec{E}=\frac{4\pi}{c^{2}}\frac{\partial }{\partial t}\vec{j},
\end{equation}
and 
\begin{equation}
\label{e2}
\nabla ^{2}\vec{B}-\frac{1}{c^{2}}\frac{\partial ^{2}}{\partial t^{2}}\vec{B}=-\frac{4\pi}{c^{2}}\nabla \times \vec{j}.
\end{equation}
Note that it is only necessary for $\vec{E}$ and $\vec{B}$ to be discussed in the far-field region, that is, Eqs. (\ref{e1}) and (\ref{e2}), which are valid in this region, are accurate, because other regions, the near-field region for example, are dominated by non-radiation fields attached to the current density \cite{kel:99} and thus are not where $\vec{E}$ and $\vec{B}$, which are supposed to propagate to the distance, can be measured.  See Ref. \cite{nov:07} for a review of near-field optics and Refs. \cite{roh:02,zan:12} for derivation of the foregoing wave equations.  No matter they are observed directly or allowed to partake in other optical processes, the emitted fields have to be determined in advance.  One basic question in electrodynamics is how to calculate the emitted fields with mathematical accuracy.  

In the literature, most theories \cite{zan:12,jac:75,zek:92} on light propagation are focused on the wave equations in Eqs. (\ref{e1}) and (\ref{e2}), especially on the operator $\nabla ^ {2}-c^{-1}\partial ^{2}/\partial t^{2}$ in these equations, because they need to use the Green's function $G$ of this operator to construct solutions to the wave equations.  In principle, $G$ has two components---$G^{(+)}$ corresponding to the retarded propagation of light and $G^{(-)}$ corresponding to the advanced propagation:
\begin{equation}
\label{e3}
G=A_{1}G^{(+)}+A_{2}G^{(-)},    
\end{equation}
where $A_{1}$ and $A_{2}$ are two constants representing the weight of each component.  Then, the assumption that the emitted fields always satisfy the principle of causality is made to determine $G$ by setting $A_{1}=1$ and $A_{2}=0$.  (It is worthwhile to note that causality was and still is a controversial notion not limited in physics \cite{dar:18}.) The result is the familiar retarded solutions to the wave equations.  

Nevertheless, as explained in Ref. \cite{guo:21}, although they are widely used in the literature, these retarded solutions can never be justified, mathematically or physically, to be the solutions to the wave equations.  It is thus fair to opine that the wave equations in (\ref{e1}) and (\ref{e2}), on which many theories on light propagation are based, are not fully solved.  One purpose of this work is to solve the wave equations in a different way, a way that does not resort to the Green function.  In the Green-function approach outlined in the preceding paragraph, it is not obvious how $A_{1}$ and $A_{2}$ can be determined in the absence of the help from the causality assumption, which, as the following discussion will show, is not well founded. 

In Ref. \cite{guo:21}, light propagation in vacuum is explained as a process involving alternate creation of electric and magnetic fields.  A magnetic field creates, in its neighborhood, an electric field (Faraday's law), and the electric field then creates another magnetic field (Amp\`{e}re's law) further away from the current density.  Such a process is a never ending process, as a result of which light propagates in vacuum.  In the light of this explanation, at each point in space, there are formally individual electric fields $\vec{E}^{(n)}$ and individual magnetic fields $\vec{B}^{(n)}$, where $n=1,2,\cdots$.  Although these individual fields do not satisfy the electromagnetic wave equations, the net fields $\vec{E}_{T}=\sum _{n=1}\vec{E}^{(n)}$ and $\vec{B}_{T}=\sum _{n=1}\vec{B}^{(n)}$ do:
\begin{equation}
\label{e4}
\nabla ^{2}\vec{E}_{T}-\frac{1}{c^{2}}\frac{\partial ^{2}}{\partial t^{2}}\vec{E}_{T}=\frac{4\pi}{c^{2}}\frac{\partial }{\partial t}\vec{j},
\end{equation}
\begin{equation}
\label{e5}
\nabla ^{2}\vec{B}_{T}-\frac{1}{c^{2}}\frac{\partial ^{2}}{\partial t^{2}}\vec{B}_{T}=-\frac{4\pi}{c^{2}}\nabla \times \vec{j}.
\end{equation}
Since the electric and magnetic fields emitted from the current density must be unique, $\vec{E}_{T}$ must be identical to the emitted electric field $\vec{E}$ in Eq. (\ref{e1}), and $\vec{B}_{T}$ must be identical to the emitted magnetic field $\vec{B}$ in Eq. (\ref{e2}).  

From the microscopic picture of light propagation, it is evident that $\vec{E}_{T}$ and $\vec{B}_{T}$ are coupled---they rely on each other to propagate.  So, while it is reasonable to remark that $\vec{E}_{T}$ and $\vec{B}_{T}$, as a whole, satisfy the principle of causality, it is questionable to assume that $\vec{E}_{T}$ and $\vec{B}_{T}$ each satisfy the principle of causality.  In this sense, the Green-function approach is not sound in physics, and, as already noted in Ref. \cite{guo:21}, the obtained retarded solutions can never be correct.

When the emitted fields are observed in experiment, they must be subject to a light-matter interaction, a process that is not covered in the wave equations, meaning that the emitted fields, which satisfy the wave equations alone, must be different from the emitted fields that are observed in experiment, and although the observed fields are causal, the emitted fields do not have to be so.  How to determine the observed fields is the second purpose of this work.

It is well known that while some optical phenomena, especially those related to light-atom interaction \cite{man:95}, such as spontaneous emission, have a straightforward explanation in quantum mechanics, others, light propagation \cite{jac:75} for example, are more conveniently discussed in classical mechanics.  The present work is devoted to light propagation and is given in the non-relativistic domain of classical physics.  See Ref. \cite{guo:07}, for example, for a quantum formulation of light propagation.

From the definitions of $\vec{E}_{T}$ and $\vec{B}_{T}$, the electromagnetic wave equations can be and are solved in Section 2 by finding those individual fields.  It turns out that the wave equations have both advanced and retarded solutions, and these solutions are equally weighted.  From the viewpoint of mathematics, the wave equations are justified in Section 2 to have such solutions.

The electromagnetic wave equations are, however, pointed out in Section 3 to have one serious limitation.  Although the emitted electric and magnetic fields are coupled through Faraday's law and Amp\`{e}re's law, the coupling is lost in the wave equations, in which the emitted fields are treated as independent or separate fields.  This limitation is discussed in Section 3 in terms of $\vec{E}^{(n)}$ and $\vec{B}^{(n)}$ and shown to be the physical reason why the emitted fields can never be causal.  So, the causality assumption in the Green-function approach must be wrong in every sense of the word.  It is also pointed out in Section 3 that the emitted fields are not those fields observed in experiment.  What is observed is other retarded fields, which are explained for their origins and are calculated in Section 3 too.  The present work is summarized in Section 4.

\section{Emitted fields}

Consider the individual electric fields $\vec{E}^{(n)}$, the first order of which $\vec{E}^{(1)}$ comes directly from the current density \cite{guo:21}:
\begin{equation}
\label{e6}
\nabla ^{2}\vec{E}^{(1)}=\frac{4\pi}{c^{2}}\frac{\partial}{\partial t}\vec{j}.
\end{equation}
In the far-field region, one particular solution of $\vec{E}^{(1)}$ reads  
\begin{eqnarray}
\label{e7}
\vec{E}^{(1)}(\vec{r},t)&=&-\frac{1}{c^{2}}\int \frac{1}{\vert \vec{r}-\vec{r}_{1}\vert }\frac{\partial }{\partial t}\vec{j}(\vec{r}_{1},t)d\vec{r}_{1}\simeq -\frac{1}{rc^{2}}\int \frac{\partial }{\partial t}\vec{j}d\vec{r}_{1}\nonumber\\
&\equiv&-\frac{1}{rc^{2}}\frac{d}{dt}j_{0}(t)\hat{z},
\end{eqnarray}
where it is understood that $d\vec{r}_{1}$ is the volume element, and $\vec{r}_{1}$ is over the volume occupied by the current density.  (The origin of the present coordinate system is assumed to reside inside the volume.) Still in the preceding equation, $r\equiv \vert \vec{r} \vert $, and the second integral on the right-hand side is assumed to point in the $\hat{z}$ direction. 

Since the current density has to be differentiable in time infinitely \cite{guo:21}, the first-order electric field $\vec{E}^{(1)}$ becomes the source of the second-order electric field $\vec{E}^{(2)}$:
\begin{equation}
\label{e8}
\nabla ^{2}\vec{E}^{(2)}=\frac{1}{c^{2}}\frac{\partial ^{2}}{\partial t^{2}}\vec{E}^{(1)}=-\frac{1}{rc^{4}}\frac{d^{3}}{dt^{3}}j_{0}(t)\hat{z}.    
\end{equation}
Evidently, $\vec{E}^{(2)}$ must be along the direction $\hat{z}$ too, that is, $\vec{E}^{(2)}=E^{(2)}\hat{z}$.  If the coordinate system is a spherical coordinate system, then Eq. (\ref{e8}) also equates
\begin{eqnarray}
\label{e9}
\frac{1}{r^{2}}\frac{\partial }{\partial r}\Big (r^{2}\frac{\partial E^{(2)}}{\partial r}\Big )&+&\frac{1}{r^{2}\sin \theta }\frac{\partial}{\partial \theta}\Big (\sin \theta \frac{\partial E^{(2)}}{\partial \theta}\Big )\nonumber\\
&+&\frac{1}{r^{2}\sin ^{2}\theta}\frac{\partial ^{2}E^{(2)}}{\partial \phi}=-\frac{1}{rc^{4}}\frac{d^{3}}{dt^{3}}j_{0}(t),
\end{eqnarray}
where $\phi$ and $\theta$ are the azimuth and polar angles respectively.  Since it has a source that depends on $r$ alone, $E^{(2)}$ should not be a function of either $\theta$ or $\phi$, meaning that Eq. (\ref{e9}) itself is simplified further to
\begin{equation}
\label{e10}
\Big ( \frac{d^{2}}{dr^{2}}+\frac{2}{r}\frac{d}{dr}\Big ) E^{(2)}=-\frac{1}{rc^{4}}\frac{d^{3}}{dt^{3}}j_{0},
\end{equation}
from which the general solutions of $E^{(2)}$ are found:
\begin{equation}
\label{e11}
E^{(2)}=-\frac{r}{2c^{4}}\frac{d^{3}}{dt^{3}}j_{0}+\frac{c_{1}}{r}+c_{2},    
\end{equation}
where $c_{1}$ and $c_{2}$ are two constants.  In the far-field region, the first term on the right-hand side of Eq. (\ref{e11}) dominates; the other two terms are all negligible.  Thus, in that region, $\vec{E}^{(2)}$ must be as follows
\begin{equation}
\label{e12}
\vec{E}^{(2)}=-\frac{r}{2c^{4}}\frac{d^{3}}{dt^{3}}j_{0}
\hat{z}.    
\end{equation}
The other individual electric fields are all similarly obtained in the far-field region.  For example,
\begin{equation}
\label{e13}
\vec{E}^{(3)}=-\frac{r^{3}}{4!c^{6}}\frac{d^{5}}{dt^{5}}j_{0}\hat{z},    
\end{equation}
and
\begin{equation}
\label{e14}
\vec{E}^{(4)}=-\frac{r^{5}}{6!c^{8}}\frac{d^{7}}{dt^{7}}j_{0}\hat{z}.    
\end{equation}

Add the individual electric fields to yield the net electric field $\vec{E}_{T}$, or the solution to the wave equation in Eq. (\ref{e4}):
\begin{eqnarray}
\label{e15}
\vec{E}_{T}&=&-\frac{1}{rc^{2}}\frac{d}{dt}\Big (j_{0}+\frac{r^{2}}{2!c^{2}}\frac{d^{2}}{dt^{2}}j_{0}+\frac{r^{4}}{4!c^{4}}\frac{d^{4}}{dt^{4}}j_{0}\nonumber\\
&&+\frac{r^{6}}{6!c^{6}}\frac{d^{6}}{dt^{6}}j_{0}+\cdots \Big )\hat{z}\nonumber\\
&=&-\frac{1}{2rc^{2}}\frac{\partial }{\partial t}\Big [ j_{0}(t+r/c)+j_{0}(t-r/c) \Big ]\hat{z}.
\end{eqnarray}

The net magnetic field $\vec{B}_{T}$, which is the solution to the magnetic wave equation (\ref{e5}), can also be obtained from the individual magnetic fields $\vec{B}^{(n)}$.  But, it turns out that $\vec{E}_{T}$ and $\vec{B}_{T}$ are not independent from each other:
\begin{equation}
\label{e16}
\nabla \times \vec{E}_{T}=-\frac{1}{c}\frac{\partial}{\partial t}\vec{B}_{T}.
\end{equation}
So, $\vec{B}_{T}$ can be calculated directly with the help of Eq. (\ref{e15}):
\begin{eqnarray}
\label{e17}
\vec{B}_{T}&=&\frac{\sin \theta}{2cr^{2}}\Big [ j_{0}(t+r/c)+j_{0}(t-r/c)\Big ]\hat{\phi}\nonumber\\
&&-\frac{\sin \theta}{2c^{2}r}\frac{\partial}{\partial t}\Big [ j_{0}(t+r/c)-j_{0}(t-r/c) \Big ]\hat{\phi}.
\end{eqnarray}
In Eq. (\ref{e17}), $\hat{\phi}$ is the unit vector along the ascending direction of $\phi$.  In the far-field region, only the second term on the right-hand side of Eq. (\ref{e17}) is needed in $\vec{B}_{T}$.  The first term, which is proportional to $1/r^{2}$ in magnitude, is only important in the other regions and is thus a non-radiation term.  In the following, the net magnetic field, when deprived of the non-radiation term, is denoted as $\vec{B}^{(R)}_{T}$:
\begin{equation}
\label{e18}
\vec{B}^{(R)}_{T}=-\frac{\sin \theta}{2c^{2}r}\frac{\partial }{\partial t}\Big [ j_{0}(t+r/c)-j_{0}(t-r/c) \Big ]
\hat{\phi}.    
\end{equation}
Since $\vec{B}_{T}$ and $\vec{E}_{T}$ are connected in Eq. (\ref{e16}), the non-radiation magnetic field in $\vec{B}_{T}$ must have a counterpart in $\vec{E}_{T}$.  A review of the calculation of $\vec{B}_{T}$ shows that the counterpart is the component of $\vec{E}_{T}$ along the $\hat{r}$ direction, a direction pointing to the ascending direction of $r$.  Such a component must be excluded from $\vec{E}_{T}$ before the radiation electric field $\vec{E}^{(R)}_{T}$ is obtained:
\begin{equation}
\label{e19}
\vec{E}^{(R)}_{T}=\frac{\sin \theta}{2rc^{2}}\frac{\partial }{\partial t} \Big [ j_{0}(t+r/c)+j_{0}(t-r/c)\Big ]
\hat{\theta},   
\end{equation}
where $\hat{\theta}$ is the unit vector along the ascending direction of $\theta$.

The present mathematical calculation confirms that, as explained mathematically and physically in Ref. \cite{guo:21}, the emitted electric field $\vec{E}^{(R)}_{T}$, or the solution to the electric wave equation, must depend on the odd-order time derivatives of the current density.  The emitted magnetic field $\vec{B}^{(R)}_{T}$, or the solution to the magnetic wave equation, must instead depend on the even-order time derivatives.  

Like other wave equations, the electromagnetic wave equations each have an advanced solution and a retarded solution; see Eqs. (\ref{e18}) and (\ref{e19}).  This result is expected mathematically, because in the far-field region the operator $\nabla ^{2}-c^{-1}\partial ^{2}/\partial t^{2}$ in these equations is approximately factored out as follows
\begin{equation}
\label{e20}
\Big (\frac{\partial}{\partial r}+\frac{1}{c}\frac{\partial}{\partial t}\Big )\Big (\frac{\partial }{\partial r}-\frac{1}{c}\frac{\partial}{\partial t}\Big ).
\end{equation}
While $\partial /\partial r-c^{-1}\partial /\partial t$ is responsible for the advanced solutions, $\partial /\partial r+c^{-1}\partial /\partial t$ is responsible for the retarded solutions.  

Equations (\ref{e18}) and (\ref{e19}) additionally show that the advanced and retarded solutions are equally important.  This observation is expected too, because in Eq. (\ref{e20}), the operators $\partial /\partial r+c^{-1}\partial /\partial t$ and $\partial /\partial r-c^{-1}\partial /\partial t$ are equally important.  In the notation of Eq. (\ref{e3}), the results in Eqs. (\ref{e18}) and (\ref{e19}) mean that $A_{1}=A_{2}=1/2$ for the electric wave equation in Eq. (\ref{e4}), and $A_{1}=-A_{2}=1/2$ for the magnetic wave equation in Eg. (\ref{e5}).

\section{Observed fields}
As far as light propagation is concerned, it is not a problem for the emitted fields $\vec{E}^{(R)}_{T}$ and $\vec{B}^{(R)}_{T}$ to have advanced components, because, in the microscopic picture of light propagation briefly discussed in Section 1, these emitted fields propagate at speed $c$ as a whole---it is in fact inappropriate to talk about the propagation of $\vec{E}^{(R)}_{T}$ or the propagation of $\vec{B}^{(R)}_{T}$.  The emitted fields do have a problem when light propagation is observed, because, doubtlessly, they do not satisfy the principle of causality \cite{roh:06}, while observed fields are known to satisfy the principle of causality.  Before the role played by the emitted fields in light observation is discussed, it is necessary to examine, from the viewpoint of physics, why the emitted fields found in Section 2 are not in conformity with causality.  The examination starts from $\vec{E}^{(n)}$ and $\vec{B}^{(n)}$, which form the emitted fields. 

It is first noted that the individual electric fields $\vec{E}^{(n)}$ are not closely related to each other, because they are only connected through the individual magnetic fields $\vec{B}^{(n)}$.  A case in point is given in Ref. \cite{guo:21}, where it is shown that $\vec{B}^{(1)}$ has to create $\vec{E}^{(1)}$ first, 
\begin{equation}
\label{e21}
\nabla \times \vec{E}^{(1)}=-\frac{1}{c}\frac{\partial }{\partial t}\vec{B}^{(1)},    
\end{equation}
and then $\vec{E}^{(1)}$ creates $\vec{B}^{(2)}$
\begin{equation}
\label{e22}
\nabla \times \vec{B}^{(2)}=\frac{1}{c}\frac{\partial}{\partial t}\vec{E}^{(1)},    
\end{equation}
and $\vec{B}^{(2)}$ creates $\vec{E}^{(2)}$
\begin{equation}
\label{e23}
\nabla \times \vec{E}^{(2)}=-\frac{1}{c}\frac{\partial }{\partial t}\vec{B}^{(2)}.    
\end{equation}
So, $\vec{E}^{(1)}$ and $\vec{E}^{(2)}$ are independent from each other in the absence of $\vec{B}^{(2)}$.  The same discussion applies to the rest individual electric fields.  When the electric wave equation, that is, Eq. (\ref{e4}), is written out in terms of $\vec{E}^{(n)}$,
\begin{equation}
\label{e24}
\nabla ^{2}\Big (\vec{E}^{(1)}+\vec{E}^{(2)}+\cdots \Big ) -\frac{1}{c^{2}}\frac{\partial ^{2}}{\partial t^{2}}\Big (\vec{E}^{(1)}+\vec{E}^{(2)}+\cdots \Big )=\frac{4\pi}{c^{2}}\frac{\partial}{\partial t}\vec{j},
\end{equation}
two points are recognized.  First, the coupling between $\vec{E}^{(n)}$ and $\vec{B}^{(n)}$ is excluded in the electric wave equation.  Second, since the coupling is excluded, the electric wave equation, as it stands, is an equation of an ensemble of independent fields.  Their independence has one consequence---these fields cannot form a process like the mutual-creation process explained briefly in Section 1 and in detail in Ref. \cite{guo:21}.  Thus, the sum of these fields, or the solution to the electric wave equation, cannot move in the vacuum.  As a result, it is not a surprise for the electric wave equation to have both advanced and retarded solutions.  See Eq. (\ref{e19}).  (The same analysis applies to the magnetic wave equation.) The emitted electric and magnetic fields can only propagate as a whole when they are coupled.  But in the wave equations (\ref{e4}) and (\ref{e5}) the emitted electric field and the emitted magnetic field are treated as independent fields.

When light is observed or detected, additional processes come into play.  (Many optical phenomena involve implicit light observation.  Light propagation in a medium composed of many particles is a multiple-scattering process \cite{guo:02} and is one such example, because in each light-scattering event the particles are driven to oscillate by the incident light, and are practically measuring the light.) Light observation or detection involves a process of light-matter interaction, in which, microscopically, charges are driven to move by an incident electric field $\vec{E}_{i}$ and an incident magnetic field $\vec{B}_{i}$ according to the Lorentz force equation
\begin{equation}
\label{e241}
\vec{F}=q\Big (\vec{E}_{i}+\frac{\vec{v}}{c}\times \vec{B}_{i}\Big ),
\end{equation}
where $\vec{F}$ is the force on a charge $q$ moving with relative velocity $\vec{v}$ to $\vec{B}_{i}$.  Usually $\vert \vec{v} \vert \ll c$, so the electric force $q\vec{E}_{i}$ is much larger than the magnetic force $q\vec{v}\times \vec{B}_{i}c^{-1}$ in magnitude, making $\vec{B}_{i}$ not as important as $\vec{E}_{i}$ in $\vec{F}$.  (The charge certainly experiences other forces, which are nevertheless ignored for their insignificance in the present discussion.  For a quantum formulation of light observation, see, for example, Ref. \cite{man:95}.) The fact that light emitted from its source is always found to satisfy the principle of causality then practically mean that, when observed, the emitted electric field from the same source, the current density in the present case, must satisfy the principle of causality.  Thus, the emitted electric field and the emitted electric field that is observed must be different.  The difference can still be understood from the Lorentz force equation.  The force $\vec{F}$ on the charge can be understood as the sum of an electric force $q\vec{E}_{i}$ and a magnetic force $q\vec{v}\times \vec{B}_{i}c^{-1}$.  Or, the same force can be understood to come from two electric fields $\vec{E}_{i}$ and $\vec{v}\times \vec{B}_{i}c^{-1}$.  In other words, if they have relative motion, then a magnetic field contributes to the net electric field observed by an observer.  This phenomenon is known in physics, for example, as the source of the R{\" o}ntgen Hamiltonian \cite{sch:01}.  

In the following, the contribution to $\vec{E}^{(R)}_{T}$ from $\vec{B}^{(R)}_{T}$ is considered through an examination of $\vec{E}^{(n)}$ and $\vec{B}^{(n)}$.  Before entering on the examination, it deserves to note that the examination should be guided by such a requirement that, aided by the contribution, the independent fields $\vec{E}^{(R)}_{T}$ should become connected and form a process similar to the mutual-creation process in Section 1, so that, as the resultant field, the observed electric field becomes a retarded field.

In the far-field region, the individual electric fields $\vec{E}^{(n)}$ are along the $\hat{\theta}$ direction, that is, $\vec{E}^{(n)}=E^{(n)}\hat{\theta}$.  The individual magnetic fields $\vec{B}^{(n)}$ are, on the other hand, along the $\hat{\phi}$ direction, meaning that it is valid to write  $\vec{B}^{(n)}=B^{(n)}\hat{\phi}$.  These observations allow, for example, $\vec{B}^{(2)}$ to be written in two ways:
\begin{equation}
\label{e25}
\vec{B}^{(2)}=B^{(2)}\hat{\phi}=\hat{r}\times \vec{E}^{(1-2)}=E^{(1-2)}\hat{\phi},    
\end{equation}
provided $\vec{E}^{(1-2)}=E^{(1-2)}\hat{\theta}$ and $E^{(1-2)}=B^{(2)}$.  It is then found that
\begin{eqnarray}
\label{e26}
\nabla \times \vec{B}^{(2)}&=&\Big (\frac{\cos \theta}{r\sin \theta}E^{(1-2)}+\frac{1}{r}\frac{\partial}{\partial \theta}E^{(1-2)}\Big )\hat{r}\nonumber\\
&&-\Big (\frac{E^{(1-2)}}{r}+\frac{\partial}{\partial r}E^{(1-2)}\Big )\hat{\theta}\nonumber\\
&\simeq&-\frac{\partial}{\partial r}E^{(1-2)}\hat{\theta},
\end{eqnarray}
where those terms whose magnitudes are suppressed by $1/r$ have been ignored.  Substitute Eq. (\ref{e26}) into (\ref{e22}) to yield
\begin{equation} 
\label{e27}
\frac{\partial E^{(1-2)}}{\partial r}=-\frac{1}{c}\frac{\partial }{\partial t}E^{(1)}.
\end{equation}
Subsequently, it is found that, still in the far-field region, $\nabla \times \vec{E}^{(2)}\simeq (\partial E^{(2)}/\partial r)\hat{\phi}$, which, when substituted into Eq. (\ref{e23}), reads
\begin{equation}
\label{e28}
\frac{\partial E^{(2)}}{\partial r}=-\frac{1}{c}\frac{\partial }{\partial t}E^{(1-2)}.
\end{equation}
Equations  (\ref{e27}) and (\ref{e28}) produce one result---the role played by $\vec{B}^{(2)}$ to connect $\vec{E}^{(1)}$ and $\vec{E}^{(2)}$, see Eqs. (\ref{e22}) and (\ref{e23}), can be taken over by $\vec{E}^{(1-2)}=\vec{B}^{(2)}\times \hat{r}$, a field formally created along the $\hat{\theta}$ direction by $\vec{B}^{(2)}$.  More specifically, $\vec{E}^{(1)}$ first formally creates $\vec{E}^{(1-2)}$, see Eq. (\ref{e27}), and $\vec{E}^{(1-2)}$ then creates, still formally, $\vec{E}^{(2)}$, see Eq. (\ref{e28}).  As such, the process 
\begin{equation}
\label{e29}
\vec{E}^{(1)}\rightarrow \vec{B}^{(2)}\rightarrow \vec{E}^{(2)}
\end{equation}
is now formally replaced by 
\begin{equation}
\label{e30}
\vec{E}^{(1)}\rightarrow \vec{E}^{(1-2)} \rightarrow \vec{E}^{(2)}.  
\end{equation}
Since, $\vec{E}^{(1-2)}$ is, like $\vec{E}^{(1)}$ and $\vec{E}^{(2)}$, along the $\hat{\theta}$ direction, it can be viewed as an effective electric field.  Note also that, according to Ref. \cite{arf:85},
\begin{eqnarray}
\label{e31}
\nabla ^{2}\vec{E}^{(2)}&=&\Big ( -\frac{2}{r^{2}}\frac{\partial E^{(2)}}{\partial \theta}-\frac{2\cos \theta}{r^{2}\sin \theta}E^{(2)}\Big )\hat{r}+\Bigg [\frac{\partial ^{2}}{\partial r^{2}}E^{(2)}\nonumber\\
&&+\frac{1}{r^{2}\sin \theta}\frac{\partial}{\partial \theta}\Big (\sin \theta \frac{\partial E^{(2)}}{\partial \theta}\Big )-\frac{E^{(2)}}{r^{2}\sin \theta}\Bigg ]\hat{\theta}\nonumber\\
&\simeq&\frac{\partial ^{2}E^{(2)}}{\partial r^{2}}\hat{\theta},
\end{eqnarray}
which shows that, in the far-field region, the relation in Eq. (\ref{e8}) becomes
\begin{equation}
\label{e32}
\frac{\partial ^{2}}{\partial r^{2}}E^{(2)}=\frac{1}{c^{2}}\frac{\partial ^{2}}{\partial r^{2}}E^{(1)}.    
\end{equation}
When Eqs.(\ref{e27}) and (\ref{e28}) are combined, a relation identical to that in the preceding equation is obtained, confirming again the validity of the introduction of $\vec{E}^{(1-2)}$.

Like $\vec{B}^{(2)}$, the rest individual magnetic fields can each formally create an effective electric field $\vec{B}^{(n)}\times \hat{r}$.  The individual electric fields $\vec{E}^{(n)}$, which are independent with each other, are now cemented by the effective electric fields to form a field chain, part of which is shown in Eq. (\ref{e30}).  In this chain, the individual electric fields and effective electric fields alternately create each other in the same way as the mutual-creation process explained in Section 1 and Ref. \cite{guo:21}.  Thus, at any point in vacuum, the total electric field $\vec{E}^{(R)}$, that is, the sum of the individual electric fields and effective electric fields, must be a propagating electric field.  Since it is composed of those fields that can alternately create each other, $\vec{E}^{(R)}$ satisfies a first-order equation, which is obtained by combining equations like (\ref{e27}) and (\ref{e28}):
\begin{eqnarray}
\label{e33}
\Big (\frac{\partial }{\partial r}+\frac{1}{c}\frac{\partial}{\partial t}\Big )\Big (E^{(1)}+E^{(1-2)}+E^{(2)}+\cdots \Big )&=&\frac{\partial }{\partial r}E^{(1)}\nonumber\\
&\simeq&0.
\end{eqnarray}
The preceding equation should be compared with Eq. (\ref{e24}) to appreciate the difference between $\vec{E}^{(R)}$ and $\vec{E}^{(R)}_{T}$.  Since the general solutions of Eq. (\ref{e33}) are a function of $r-ct$, the total field must be in addition a retarded wave.  In its vector form, $\vec{E}^{(R)}$ is indeed found to be retarded:
\begin{equation}
\label{e34}
\vec{E}^{(R)}=\vec{E}^{(R)}_{T}+\vec{B}^{(R)}_{T}\times\hat{r}=\frac{\sin \theta}{c^{2}r}\frac{\partial }{\partial t}j_{0}(t-r/c)\hat{\theta}.
\end{equation}
The field $\vec{E}^{(R)}$ is nothing other than the observed electric field at $\vec{r}$ when light emitted from the current density is observed.

The magnetic field $\vec{B}^{(R)}$ observed by the same observer should be likewise obtained from not only $\vec{B}^{(R)}_{T}$ but also a net effective magnetic field $\hat{r}\times \vec{E}^{(R)}_{T}$, that is,
\begin{equation}
\label{e35}
\vec{B}^{(R)}=\vec{B}^{(R)}_{T}+\hat{r}\times \vec{E}^{(R)}_{T}=\frac{\sin \theta}{rc^{2}}\frac{\partial}{\partial \theta }j_{0}(t-r/c)\hat{\phi}.    
\end{equation}

When light emitted from the current density is not observed, the emitted electric field $\vec{E}^{(R)}_{T}$ and emitted magnetic field $\vec{B}^{(R)}_{T}$ alternately create each other and propagate as a whole at speed $c$ in vacuum.  When the light is observed, that is, when both electric and magnetic fields are observed, the observed electric field is not just the emitted electric field and should also include an effective electric field coming from the emitted magnetic field.  Similarly, the observed magnetic field should have a contribution from the emitted electric field too.  Both the observed electric field and the observed magnetic field must be retarded fields.  Thus, whether the emitted light is observed or not, the principle of causality is always preserved.  Causality does not need to be preserved by hand \cite{roh:02} in classical physics.

That the electromagnetic wave equations have advanced and retarded solutions is well known in the literature and used as evidence by some authors \cite{sch:70} to criticize that Maxwell's equations, from which the wave equations are derived, are incomplete.  Contrary to the criticism, the discussion in this work demonstrates that Maxwell's equations are compete in every sense of the word.  The criticism cannot be justified, because it misses at least two points, that is, the electromagnetic wave equations are unable to describe fully light propagation and the emitted fields and observed fields are unequal. 

The observed fields $\vec{E}^{(R)}$ and $\vec{B}^{(R)}$ are in fact identical to $\vec{E}^{(R)}_{T}$ and $\vec{B}^{(R)}_{T}$ respectively, when the advanced components of the latter fields are dropped and the retarded components are doubled in magnitude.  But, such a result should be viewed at most as a coincidence and should not be used to justify the conventional Green-function method outlined in Section 1.  The Green-function approach is fundamentally wrong, because it is built on a simplified explanation of light propagation and on false assumptions.  See Section 1.

\section{Conclusion}

Light propagation and observed light propagation are two different processes, each of which is a representation of the mathematics and physics that have evaded detailed analysis.  After the mathematics and physics are analyzed, the said processes become clear, and some questions, such as if classical physics encompasses causality and if Maxwell equations are complete, are answered too.

\end{document}